# Coherence and phase synchronization: generalization to pairs of multivariate time series, and removal of zero-lag contributions

Roberto D. Pascual-Marqui

The KEY Institute for Brain-Mind Research
University Hospital of Psychiatry
Lenggstr. 31, CH-8032 Zurich, Switzerland
pascualm <at> key.uzh.ch
www.keyinst.uzh.ch/loreta

## *Abstract*

Coherence and phase synchronization between time series corresponding to different spatial locations are usually interpreted as indicators of the "connectivity" between locations. In neurophysiology, time series of electric neuronal activity are essential for studying brain interconnectivity. Such signals can either be invasively measured from depth electrodes, or computed from very high time resolution, non-invasive, extracranial recordings of scalp electric potential differences (EEG: electroencephalogram) and magnetic fields (MEG: magnetoencephalogram) by means of a tomography such as sLORETA (standardized low resolution brain electromagnetic tomography). There are two problems in this case. First, in the usual situation of unknown cortical geometry, the estimated signal at each brain location is a vector with three components (i.e. a current density vector), which means that coherence and phase synchronization must be generalized to pairs of multivariate time series. Second, the inherent low spatial resolution of the EEG/MEG tomography introduces artificially high zero-lag coherence and phase synchronization. In this report, solutions to both problems are presented. Two additional generalizations are briefly mentioned: (1) conditional coherence and phase synchronization; and (2) non-stationary time-frequency analysis. Finally, a non-parametric randomization method for connectivity significance testing is outlined. The new connectivity measures proposed here can be applied to pairs of univariate EEG/MEG signals, as is traditional in the published literature. However, these calculations cannot be interpreted as "connectivity", since it is in general incorrect to associate an extracranial electrode or sensor to the underlying cortex.

## *Notation and definitions*

The terms "multivariate time series", "multiple time series", and "vector time series" have identical meaning in this paper.

For general notation and definitions, see e.g. Brillinger (1981) for stationary multivariate time series analysis, and see e.g. Mardia et al (1979) for general multivariate statistics.

Let $\mathbf{X}_{jt} \in \mathbb{R}^{p \times 1}$ and $\mathbf{Y}_{jt} \in \mathbb{R}^{q \times 1}$ denote two stationary multivariate time series, for discrete time $t = 0 \ldots N_T - 1$, with $j = 1 \ldots N_R$ denoting the $j$-th time segment. The discrete Fourier transforms are denoted as $\mathbf{X}_{j\omega} \in \mathbb{C}^{p \times 1}$ and $\mathbf{Y}_{j\omega} \in \mathbb{C}^{q \times 1}$, and defined as:

Eq. 1: $\quad \mathbf{X}_{j\omega} = \sum_{t=0}^{N_T - 1} \mathbf{X}_{jt} e^{-2\pi i \omega t / N_T}$

Eq. 2: $\quad \mathbf{Y}_{j\omega} = \sum_{t=0}^{N_T - 1} \mathbf{Y}_{jt} e^{-2\pi i \omega t / N_T}$

for discrete frequencies $\omega = 0 \ldots N_T - 1$, and where $i = \sqrt{-1}$.

It will be assumed throughout that $\mathbf{X}_\omega$ and $\mathbf{Y}_\omega$ each have zero mean.

Let:

Eq. 3: $\quad \mathbf{S}_{\mathbf{XX}\omega} = \frac{1}{N_R} \sum_{j=1}^{N_R} \mathbf{X}_{j\omega} \mathbf{X}_{j\omega}^*$

Eq. 4: $\quad \mathbf{S}_{\mathbf{YY}\omega} = \frac{1}{N_R} \sum_{j=1}^{N_R} \mathbf{Y}_{j\omega} \mathbf{Y}_{j\omega}^*$

Eq. 5: $\quad \mathbf{S}_{\mathbf{XY}\omega} = \frac{1}{N_R} \sum_{j=1}^{N_R} \mathbf{X}_{j\omega} \mathbf{Y}_{j\omega}^*$

Eq. 6: $\quad \mathbf{S}_{\mathbf{YX}\omega} = \mathbf{S}_{\mathbf{XY}\omega}^* = \frac{1}{N_R} \sum_{j=1}^{N_R} \mathbf{Y}_{j\omega} \mathbf{X}_{j\omega}^*$

denote complex valued covariance matrices, where the superscript "*" denotes vector/matrix transposition and complex conjugation. Note that $\mathbf{S}_{\mathbf{XX}\omega}$ and $\mathbf{S}_{\mathbf{YY}\omega}$ are Hermitian matrices, satisfying $\mathbf{S} = \mathbf{S}^*$. When multiplied by the factor $(2\pi N_T)^{-1}$, these matrices correspond to the cross-spectral density matrices.

## *Coherence between pairs of multivariate time series*

In the case of real-valued stochastic variables, Mardia et al (1979) review several "measures of correlation between vectors". These definitions can be straightforwardly generalized to the complex valued domain. In particular, this work will make use of a general measure of correlation proposed by Kent (1983), which is closely related to the vector alienation coefficient (Hotelling 1936, Mardia et al 1979). This measure of general coherence is also equivalent to the coefficient of determination as defined by Pierce (1982).

The general coherence $\rho_G$ is defined as:

Eq. 7: $\quad \rho_G^2 = 1 - \frac{\left| \mathbf{S}_{\mathbf{YY}\omega / \mathbf{X}} \right|}{\left| \mathbf{S}_{\mathbf{YY}\omega} \right|}$

where:

Eq. 8: $\quad \mathbf{S}_{\mathbf{YY}\omega / \mathbf{X}} = \mathbf{S}_{\mathbf{YY}\omega} - \mathbf{S}_{\mathbf{YX}\omega} \mathbf{S}_{\mathbf{XX}\omega}^{-1} \mathbf{S}_{\mathbf{XY}\omega}$

denotes the conditional variance of $\mathbf{Y}_\omega$ given $\mathbf{X}_\omega$.

This is a multivariate generalization of the ordinary squared correlation coefficient between two real-valued univariate stochastic variables. In addition, for the case of two complex-valued univariate stochastic variables, it is the ordinary squared coherence (see e.g. Equation 3 in Nolte et al 2004).

Note that the general coherence can equivalently be defined as:

Eq. 9: $\rho_G^2 = 1 - \frac{|\mathbf{S}_{XX\omega/Y}|}{|\mathbf{S}_{XX\omega}|}$

This means the it is a symmetric measure of association between $\mathbf{X}_\omega$ and $\mathbf{Y}_\omega$.

The general coherence takes values in the range zero (in the case of linear independence between $\mathbf{X}_\omega$ and $\mathbf{Y}_\omega$) to one (in the case of perfect linear prediction between $\mathbf{X}_\omega$ and $\mathbf{Y}_\omega$).

## *Phase synchronization between pairs of univariate time series*

The term "phase synchronization" has a very rigorous physics definition (see e.g. Rosenblum et al 1996). The basic idea behind this definition has been adapted and used to great advantage in the neurosciences (Tass et al 1998, Quian-Quiroga et al 2002, Pereda et al 2005, Stam et al 2007), as in for example, the analysis of pairs of time series of measured scalp electric potentials differences (i.e. EEG: electroencephalogram). Other equivalent descriptive names for "phase synchronization" that appear in the neurosciences are phase locking, phase locking value, phase locking index, phase coherence, and so on.

An informal definition for the statistical "phase synchronization" model will now be given. In order to simplify this informal definition even further, it will be assumed that the two univariate time series (i.e. $p = q = 1$) of interest are stationary. At a given discrete frequency $\omega$, the sample data in the frequency domain (using the discrete Fourier transform) is denoted as $x_{j\omega}, y_{j\omega} \in \mathbb{C}$, with $j = 1...N_R$ denoting the $j$-th time segment. If the phase difference $\Delta\varphi_j = \varphi_j^x - \varphi_j^y$ is "stable" over time segments $j$, regardless of the amplitudes, then there is a "connection" between the locations at which the measurements were made. A measure of stability of phase difference is precisely "phase synchronization". It can as well be defined for the non-stationary case, using concepts of time-varying instantaneous phase, and defining stability over time (instead of stability over time segments).

In the case of univariate time series, i.e. $p = q = 1$, phase synchronization can be viewed as the modulus (absolute value) of the complex valued (Hermitian) coherency between the *normalized* Fourier transforms. This interpretation will be the basis for a generalization to the multivariate case later on.

These variables are normalized prior to the coherency calculation in order to remove from the outset any amplitude effect, leaving only phase information. This normalization operation is highly non-linear.

The modulus of the coherency is used as a measure for phase synchronization because it is conveniently bounded in the range zero (no synchronization) to one (perfect synchronization).

Formally, let $x_{j\omega}, y_{j\omega} \in \mathbb{C}$ denote the discrete Fourier transforms of the univariate time series. The normalized variables correspond to $\breve{x}_{j\omega} = \frac{x_{j\omega}}{|x_{j\omega}|}$ and $\breve{y}_{j\omega} = \frac{y_{j\omega}}{|y_{j\omega}|}$. Note that, by definition, their real-valued Hermitian variances take the value 1:

Eq. 10: $s_{\breve{x}\breve{x}\omega} = \frac{1}{N_R}\sum_{j=1}^{N_R} \breve{x}_{j\omega}\breve{x}_{j\omega}^* = \frac{1}{N_R}\sum_{j=1}^{N_R} \frac{x_{j\omega}}{|x_{j\omega}|}\frac{x_{j\omega}^*}{|x_{j\omega}|} = 1$

Eq. 11: $s_{\breve{y}\breve{y}\omega} = \frac{1}{N_R}\sum_{j=1}^{N_R} \breve{y}_{j\omega}\breve{y}_{j\omega}^* = \frac{1}{N_R}\sum_{j=1}^{N_R} \frac{y_{j\omega}}{|y_{j\omega}|}\frac{y_{j\omega}^*}{|y_{j\omega}|} = 1$

Therefore, the modulus of the complex valued (Hermitian) coherency between these <u>normalized</u> variables is simply:

Eq. 12: $PS = \left|\frac{s_{\breve{x}\breve{y}\omega}}{\sqrt{s_{\breve{x}\breve{x}\omega}s_{\breve{y}\breve{y}\omega}}}\right| = |s_{\breve{x}\breve{y}\omega}| = \left|\frac{1}{N_R}\sum_{j=1}^{N_R} \breve{x}_{j\omega}\breve{y}_{j\omega}^*\right| = \sqrt{\left[\text{Re}(s_{\breve{x}\breve{y}\omega})\right]^2 + \left[\text{Im}(s_{\breve{x}\breve{y}\omega})\right]^2}$

where $\text{Re}(s_{\breve{x}\breve{y}\omega})$ and $\text{Im}(s_{\breve{x}\breve{y}\omega})$ denote the real and imaginary parts of $s_{\breve{x}\breve{y}\omega}$.

## *Phase synchronization between pairs of multivariate time series*

Based on the foregoing arguments, a natural definition for phase synchronization between two vector time series follows: it is the general coherence between the normalized vectors.

There are at least two possibilities for defining normalization.

Vector-wise normalization is:

Eq. 13: $\begin{cases} \breve{\mathbf{X}}_{j\omega} = (\mathbf{X}_{j\omega}^*\mathbf{X}_{j\omega})^{-1/2}\mathbf{X}_{j\omega} \\ \breve{\mathbf{Y}}_{j\omega} = (\mathbf{Y}_{j\omega}^*\mathbf{Y}_{j\omega})^{-1/2}\mathbf{Y}_{j\omega} \end{cases}$

Variable-wise normalization is:

Eq. 14: $\begin{cases} \breve{\mathbf{X}}_{j\omega} = [diag(\mathbf{X}_{j\omega}\mathbf{X}_{j\omega}^*)]^{-1/2}\mathbf{X}_{j\omega} \\ \breve{\mathbf{Y}}_{j\omega} = [diag(\mathbf{Y}_{j\omega}\mathbf{Y}_{j\omega}^*)]^{-1/2}\mathbf{Y}_{j\omega} \end{cases}$

where the "*diag*" matrix operator sets to zero the off-diagonal elements, thus creating the corresponding diagonal matrix.

Once a certain form of normalization is chosen (either Eq. 13 or Eq. 14), the normalized variables are used for computing the covariances as in Eq. 3 to Eq. 6, and finally plugging them into Eq. 7 to Eq. 8, to give phase synchronization.

Formally, the general phase synchronization is defined as:

Eq. 15: $PS_G = \rho_G = \sqrt{1 - \frac{|\mathbf{S}_{\breve{Y}\breve{Y}\omega/\breve{X}}|}{|\mathbf{S}_{\breve{Y}\breve{Y}\omega}|}} = \sqrt{1 - \frac{|\mathbf{S}_{\breve{Y}\breve{Y}\omega} - \mathbf{S}_{\breve{Y}\breve{X}\omega}\mathbf{S}_{\breve{X}\breve{X}\omega}^{-1}\mathbf{S}_{\breve{X}\breve{Y}\omega}|}{|\mathbf{S}_{\breve{Y}\breve{Y}\omega}|}}$

Note that this generalization reduces to the classical definition for univariate time series (see Eq. 12), regardless of which normalization form (Eq. 13 or Eq. 14) is used.

## *Zero-lag contribution to coherence and phase synchronization: problem description*

In some fields of application, the coherence or phase synchronization between two time series corresponding to two different spatial locations is interpreted as a measure of the "connectivity" between those two locations.

For example, consider the time series of scalp electric potential differences (EEG: electroencephalogram) at two locations. The coherence or phase synchronization is interpreted by some researchers as a measure of "connectivity" between the underlying cortices (see e.g. Nolte et al 2004 and Stam et al 2007).

However, even if the underlying cortices are not actually connected, significantly high coherence or phase synchronization might still occur due to the volume conduction effect: activity at any cortical area will be observed instantaneously (zero-lag) by all scalp electrodes.

As a possible solution to this problem, the electric neuronal activity distributed throughout the cortex can be estimated from the EEG by using imaging techniques such as standardized low resolution brain electromagnetic tomography (sLORETA) (Pascual-Marqui et al 2002). At each voxel in the cortical grey matter, a 3-component vector time series is computed, corresponding to the current density vector with dipole moments along axes *X*, *Y*, and *Z*. This tomography has the unique properties of being linear, of having zero localization error, but of having low spatial resolution. Due to such spatial "blurring", the time series will again suffer from non-physiological inflated values of zero-lag coherence and phase synchronization.

Formally, consider two different spatial locations where there is no actual activity. However, due to a third truly active location, and because of low spatial resolution (or volume conductor type effect), there is some measured activity at these locations:

Eq. 16: $$\begin{cases} \mathbf{X}_{jt} = \mathbf{C}\mathbf{Z}_{jt} + \boldsymbol{\varepsilon}_{jt}^{x} \\ \mathbf{Y}_{jt} = \mathbf{D}\mathbf{Z}_{jt} + \boldsymbol{\varepsilon}_{jt}^{y} \end{cases}$$

where $\mathbf{Z}_{jt}$ is the time series of the truly active location; $\mathbf{C}$ and $\mathbf{D}$ are matrices determined by the properties of the low spatial resolution problem; and $\boldsymbol{\varepsilon}_{jt}^{x}$ and $\boldsymbol{\varepsilon}_{jt}^{y}$ are independent and identically distributed random white noise.

In this model, although **X** and **Y** are not "connected", coherence and phase synchronization will indicate some connection, due to zero-lag spatial blurring.

Things can get even worse due to the zero-lag effect. Suppose that two time series are measured under two different conditions in which the zero-lag blurring effect is constant. The goal is to perform a statistical test to compare if there is a change in connectivity. Since the zero-lag effect is the same in both conditions, then it should seemingly not account for any significant difference in coherence or phase synchronization. However, this might be very misleading. In the model in Eq. 16, a simple increase in the signal to noise ratio (e.g. by increasing the norms of **C** and **D**) will produce an increase in coherence and phase synchronization, due again to the zero-lag effect. This example shows that the zero-lag effect can render meaningless a comparison of two or more conditions.

Recently, two proposals have been published to obtain estimates of lagged connectivity, which cannot be influenced by the zero-lag blurring (or volume conductor) effect. One proposal is the use of the imaginary part of the coherency (Nolte et al 2004), because the real part is the one mostly affected by the zero-lag effect. The second proposal, termed the phase lag index, consists of computing phase synchronization in such a way that it not be influenced by phase differences that center around 0 mod $\pi$ (Stam et al 2007).

Here we propose a direct solution: to partial out (i.e. to remove) the zero-lag instantaneous interactions, and to compute coherence and phase synchronization using the residual, corrected time series. Such a method seems to be totally justified: for example, in a seminal paper on linear feedback by Geweke (1982), this type of approach was used to define several measures of causal interactions between time series.

The method presented here is general and can be used on any multiple time series, regardless of their nature.

We wish to emphasize that we will very explicitly not apply the methods developed here to the measured time series of scalp electric potentials (EEG) nor to magnetic fields (MEG). The new methods are intended for use with estimates of electric neuronal activity, by means of a tomography that has been validated, in theory and in practice, such as sLORETA (Pascual-Marqui 2002). The reason for this clarification and recommendation is because the direct visual inspection of scalp EEG and MEG can be misleading as a tool for the localization of brain activity and interconnectivity: it is simply naïve and incorrect to associate an electrode or sensor to the underlying cortex. Actually, the laws of electrodynamics that relate sources with electric potentials and magnetic fields are fairly complicated and do not justify such a method for brain localization inference. This should be taken into account when interpreting the results of many publications with wire-diagrams based on significant connections between scalp electrode time series: these extracranial-based wires do not necessarily correspond to "wires" connecting the underlying cortices.

### *Identifying the zero-lag contribution*

The zero-lag contribution to Hermitian covariances between time series must first be estimated in order to partial out (remove) its effect.

Let $\mathbf{Z}_{jt} \in \mathbb{R}^{r \times 1}$, for $t = 0 \ldots N_T - 1$ and for $j = 1 \ldots N_R$, denote a multivariate time series. Let $\mathbf{Z}_{j\omega} \in \mathbb{C}^{r \times 1}$ denote its discrete Fourier transform, at discrete frequency $\omega$, for $\omega = 0 \ldots N_T - 1$. Let $\mathbf{S}_{\mathbf{ZZ}\omega}$ denote its Hermitian covariance matrix (see e.g. Eq. 3).

The zero-lag contribution to the Hermitian covariance can be estimated in the following steps:

1. Take the original time series $\mathbf{Z}_{jt}$ and filter it to leave exclusively the frequency $\omega$ component. Denote the filtered time series as $\left(\mathbf{Z}_{jt}^{\omega-Filtered}\right)$. Note that, by construction, the spectral density of $\left(\mathbf{Z}_{jt}^{\omega-Filtered}\right)$ is zero everywhere except at frequency $\omega$.

2. Compute the real-valued, zero-lag, time domain, symmetric covariance matrix for the filtered time series $\left(\mathbf{Z}_{jt}^{\omega-Filtered}\right)$ at frequency $\omega$:

Eq. 17: $\quad \mathbf{A}_\omega = \dfrac{1}{N_T N_R} \sum_{j=1}^{N_R} \sum_{t=1}^{N_t} \left(\mathbf{Z}_{jt}^{\omega-Filtered}\right)\left(\mathbf{Z}_{jt}^{\omega-Filtered}\right)^T \in \mathbb{R}^{r \times r}$

Making use of Parseval's theorem for the filtered time series, the following relation holds:

Eq. 18: $\quad \operatorname{Re}\left(\mathbf{S}_{ZZ\omega}\right) = \dfrac{N_T^2}{2} \mathbf{A}_\omega$

where $\operatorname{Re}\left(\mathbf{S}_{ZZ\omega}\right)$ denotes the real part of $\mathbf{S}_{ZZ\omega}$.

Therefore, the zero-lag contribution to multivariate covariances is contained in the real part of the Hermitian covariance. This basic result will allow a rigorous solution to the elimination of the zero-lag confounding effect. The details follow.

## *The "zero-lag removed" coherence between pairs of multivariate time series*

Define the joint time series:

Eq. 19: $\quad \mathbf{Z}_{jt} = \begin{pmatrix} \mathbf{Y}_{jt} \\ \mathbf{X}_{jt} \end{pmatrix}$

with discrete Fourier transform:

Eq. 20: $\quad \mathbf{Z}_{j\omega} = \begin{pmatrix} \mathbf{Y}_{j\omega} \\ \mathbf{X}_{j\omega} \end{pmatrix}$

and Hermitian covariance:

Eq. 21: $\quad \mathbf{S}_{ZZ\omega} = \begin{pmatrix} \mathbf{S}_{YY\omega} & \mathbf{S}_{YX\omega} \\ \mathbf{S}_{XY\omega} & \mathbf{S}_{XX\omega} \end{pmatrix}$

Let $\operatorname{Re}\left(\mathbf{S}_{ZZ\omega}\right)$ denote the real part of $\mathbf{S}_{ZZ\omega}$.

The general lagged coherence $\rho_{GL}$ (i.e. the zero-lag removed general coherence) is defined as:

Eq. 22: $\quad \rho_{GL}^2 = 1 - \dfrac{\left|\mathbf{S}_{ZZ\omega}\right|}{\left|\operatorname{Re}\left(\mathbf{S}_{ZZ\omega}\right)\right|}$

A detailed analysis of the classical case of two univariate time series ($x_{jt}$ and $y_{jt}$) will serve to illustrate this new measure of general lagged coherence. Let $s_{yy\omega}$ and $s_{xx\omega}$ denote the pure real variances, $s_{xy\omega}$ the complex valued covariance, with real and imaginary parts denoted as $\text{Re}(s_{xy\omega})$ and $\text{Im}(s_{xy\omega})$. Then Eq. 22 is:

Eq. 23: $$\rho_{GL}^2 = 1 - \frac{|\mathbf{S}_{ZZ\omega}|}{|\text{Re}(\mathbf{S}_{ZZ\omega})|} = 1 - \frac{\begin{vmatrix} s_{yy\omega} & s_{yx\omega} \\ s_{xy\omega} & s_{xx\omega} \end{vmatrix}}{\begin{vmatrix} s_{yy\omega} & \text{Re}(s_{xy\omega}) \\ \text{Re}(s_{xy\omega}) & s_{xx\omega} \end{vmatrix}}$$

which can be written as:

Eq. 24: $$\rho_{GL}^2 = 1 - \frac{s_{yy\omega} - \left\{[\text{Re}(s_{xy\omega})]^2 + [\text{Im}(s_{xy\omega})]^2\right\}/s_{xx\omega}}{s_{yy\omega} - [\text{Re}(s_{xy\omega})]^2/s_{xx\omega}}$$

As seen from Eq. 24, the general lagged coherence contains a quotient of residual variances. The denominator corresponds to the variance of time series $y$ conditional on the real part. The numerator corresponds to the variance of time series $y$ conditional on the real and imaginary parts. This is precisely what was required: a rigorous coherence measure where the lag-zero contribution is partialled out.

In other words: the general lagged coherence is defined as the partial coherence between the complex-valued stochastic variables $(x_{j\omega}, y_{j\omega})$, with the zero-lag effect removed.

Eq. 24 can be equivalently written as:

Eq. 25: $$\rho_{GL}^2 = \frac{[\text{Im}(s_{yx\omega})]^2}{s_{yy\omega}s_{xx\omega} - [\text{Re}(s_{yx\omega})]^2}$$

Its signed square root is:

Eq. 26: $$\rho_{GL} = \frac{\text{Im}(s_{yx\omega})}{\sqrt{s_{yy\omega}s_{xx\omega} - [\text{Re}(s_{yx\omega})]^2}}$$

The new result in Eq. 26 is quite different from the proposed version by Nolte et al (2004) that consists of the imaginary part of the coherency:

Eq. 27: $$\text{Im}(Coherency) = \frac{\text{Im}(s_{yx\omega})}{\sqrt{s_{yy\omega}s_{xx\omega}}}$$

### *The "zero-lag removed" phase synchronization between multivariate time series*

Based on Eq. 15 and Eq. 22, the general lagged phase synchronization (i.e. the "zero-lag removed" general phase synchronization) between multivariate time series **X** and **Y** is defined as:

Eq. 28: $PS_{GL} = \rho_{GL} = \sqrt{1 - \frac{|\mathbf{S}_{\breve{Z}\breve{Z}\omega}|}{|\text{Re}(\mathbf{S}_{\breve{Z}\breve{Z}\omega})|}}$

where:

Eq. 29: $\mathbf{S}_{\breve{Z}\breve{Z}\omega} = \frac{1}{N_R} \sum_{j=1}^{N_R} \breve{\mathbf{Z}}_{j\omega} \breve{\mathbf{Z}}_{j\omega}^*$

In Eq. 29, $\breve{\mathbf{Z}}_{j\omega}$ denotes the normalized joint time series:

Eq. 30: $\breve{\mathbf{Z}}_{j\omega} = \begin{pmatrix} \breve{\mathbf{Y}}_{j\omega} \\ \breve{\mathbf{X}}_{j\omega} \end{pmatrix}$

Recall that two normalization variants exist for $\breve{\mathbf{Y}}_{j\omega}$ and $\breve{\mathbf{X}}_{j\omega}$, as defined above in Eq. 13 and Eq. 14.

A detailed analysis of the classical case of two univariate time series ($x_{jt}$ and $y_{jt}$) will serve to illustrate this new measure of general lagged phase synchronization. Note that these results are unique and independent of the form of normalization (Eq. 13 or Eq. 14). The variances of the normalized variables are equal to one, i.e. $s_{\breve{y}\breve{y}\omega} = s_{\breve{x}\breve{x}\omega} = 1$, as demonstrated in Eq. 10 and Eq. 11 above. Let $s_{\breve{x}\breve{y}\omega}$ the complex valued covariance for the normalized variables, with real and imaginary parts denoted as $\text{Re}(s_{\breve{x}\breve{y}\omega})$ and $\text{Im}(s_{\breve{x}\breve{y}\omega})$. Then Eq. 28 is:

Eq. 31: $PS_{GL} = \rho_{GL} = \sqrt{1 - \frac{|\mathbf{S}_{\breve{Z}\breve{Z}\omega}|}{|\text{Re}(\mathbf{S}_{\breve{Z}\breve{Z}\omega})|}} = \sqrt{1 - \frac{\begin{vmatrix} 1 & s_{\breve{y}\breve{x}\omega} \\ s_{\breve{x}\breve{y}\omega} & 1 \end{vmatrix}}{\begin{vmatrix} 1 & \text{Re}(s_{\breve{y}\breve{x}\omega}) \\ \text{Re}(s_{\breve{x}\breve{y}\omega}) & 1 \end{vmatrix}}}$

which can be written as:

Eq. 32: $PS_{GL} = \rho_{GL} = \sqrt{1 - \frac{1 - \left\{[\text{Re}(s_{\breve{x}\breve{y}\omega})]^2 + [\text{Im}(s_{\breve{x}\breve{y}\omega})]^2\right\}}{1 - [\text{Re}(s_{\breve{x}\breve{y}\omega})]^2}}$

As seen from Eq. 32, the general lagged phase synchronization contains a quotient of residual variances. The denominator corresponds to the variance of (normalized) time series $y$ conditional on the real part. The numerator corresponds to the variance of (normalized) time series $y$ conditional on the real and imaginary parts. This is precisely what was required: a rigorous phase synchronization measure where the lag-zero contribution is partialled out.

In other words: the general lagged phase synchronization is defined as the partial coherence between the normalized complex-valued stochastic variables $(\breve{x}_{j\omega}, \breve{y}_{j\omega})$, with the zero-lag effect removed.

Eq. 32 can be equivalently written as:

Eq. 33: $PS_{GL} = \sqrt{\dfrac{\left[\text{Im}(s_{\tilde{x}\tilde{y}\omega})\right]^2}{1-\left[\text{Re}(s_{\tilde{x}\tilde{y}\omega})\right]^2}}$

The new result in Eq. 33 is quite different from the classical phase synchronization measure (see Eq. 12 above). In addition, the new result in Eq. 33 is quite different from the phase lag index (Stam et al 2007):

Eq. 34: $PLI = \left| \dfrac{1}{N_R} \sum_{j=1}^{N_R} sign\left[\text{Im}(y_{j\omega} x^*_{j\omega})\right] \right|$

## *Some further generalizations: time-frequency methods*

All the previous development can be applied to any form of time-varying Fourier transforms or wavelets, such as, for instance, Gabor or Morlet transforms. In this case, the subscript "j" now would refer to time, in contrast to its previous interpretation in the stationary case in which it indexed "time segment".

## *Some further generalizations: conditional coherence and phase synchronization*

Consider the case where in addition to the two time series of interest **X** and **Y**, there is now a third one **Z**. It is relatively straightforward to now extend the definitions of coherence and phase synchronization (with or without zero-lag removal) for **X** and **Y** conditional on **Z**.

Examples of related work on conditional measures can be found in Geweke (1984) and Schelter et al (2006).

## *Non-parametric randomization tests for coherence and phase synchronization*

The statistical methodology referred to in this section, can be studied in detail in Manly (1997) and in Nichols and Holmes (2001).

As a particular case, consider the general coherence $\rho_G^2$ for two multiple time series with data pairs $(\mathbf{X}_{j\omega}, \mathbf{Y}_{j\omega})$, for $j=1...N_R$. The results that follow can be applied in identical fashion to any of the connectivity measures described above (phase synchronization, general phase synchronization, general lagged coherence, general lagged phase synchronization, imaginary part of coherency, and phase lag index).

The question of interest is to test the null hypothesis that the coherence is zero.

For this purpose, we estimate the probability distribution under the null hypothesis, i.e. under conditions of zero coherence. This is performed by destroying any possible actual coherence, via randomization of subscript "j" for one of the time series (e.g. $\mathbf{X}_{j\omega}$), and recomputing the coherence. For instance, the actual data is:

$$(\mathbf{X}_{1\omega}, \mathbf{Y}_{1\omega}), (\mathbf{X}_{2\omega}, \mathbf{Y}_{2\omega}), (\mathbf{X}_{3\omega}, \mathbf{Y}_{3\omega}), ....(\mathbf{X}_{N_R\omega}, \mathbf{Y}_{N_R\omega})$$

and a randomized sample corresponding to the null hypothesis (zero coherence) might lead to the new pairs:

$$(\mathbf{X}_{5\omega}, \mathbf{Y}_{1\omega}), (\mathbf{X}_{9\omega}, \mathbf{Y}_{2\omega}), (\mathbf{X}_{1\omega}, \mathbf{Y}_{3\omega}), ....(\mathbf{X}_{2\omega}, \mathbf{Y}_{N_R\omega})$$

The collection of randomized coherences (after many randomizations) allows estimation of the empirical probability distribution against which the actual coherence is compared.

This same methodology can be applied to all other measures defined in this paper (e.g. phase synchronization, with or without zero-lag correction).

Another example of randomization tests for phase synchronization, which is related but not identical to the method presented here, can be found in Allefeld and Kurths (2004).

## *Acknowledgements*

I have had extremely useful discussions with G. Nolte, C. Allefeld, and J. Wackermann.

G. Nolte pointed out a number of embarrassing inconsistencies I wrote in the first draft of this technical report, regarding the correction for zero-lag effect. Thanks to his comments, I came up with the current definition for partialling out the zero-lag effect.

C. Allefeld made a number of useful comments on the rigorous physics definition of phase synchronization, which I tried to keep in mind while explaining in an informal manner how this concept has been employed as a statistical measure of connectivity in the neurosciences.

J. Wackermann made a number of useful comments on my original informal and mostly incorrect use of statistics terminology. I tried to correct these problems.